\documentclass[12pt,a4paper]{article}
\usepackage[latin1]{inputenc}
\usepackage{amsmath}
\usepackage{amsfonts}
\usepackage{amssymb}
\usepackage{graphicx}

\begin{document}

\date{}

\author{ S. K.Tripathy\footnote{Department of Physics,
Indira Gandhi Institute of Technology,
Sarang, Dhenkanal, Odisha-759146, INDIA,
tripathy\_ sunil@rediffmail.com}}

\title  {\bf  Refractive Indices of Semiconductors from Energy gaps}

\maketitle

\begin{abstract}
An empirical relation based on energy gap and refractive index data has been proposed in the present study to calculate the refractive index of semiconductors. The proposed model is then applied to binary as well as ternary semiconductors for a wide range of energy gap. Using the relation, dielectric constants of some III-V group semiconductors are calculated. The calculated values  for different group of binary semiconductors, alkali halides and ternary semiconductors fairly agree with other calculations and known values over a wide range of energy gap. The temperature variation of refractive index for some binary semiconductors have been calculated.
\end{abstract}


\section{Introduction}

With the advent of recent technologies, novel semiconductors rich in their optoelectronic properties find wide range of  applications in optical, electronic and optoelectronic devices such as light emitting diodes (LED), laser diodes (LD), integrated circuits (IC), photo detectors (PD), nanotechnology, heterostructure lasers and optical modulators operating in mid infra-red regions ($2-5 \mu m$) \cite{Paskov97,Rappl01}. The optical and electronic behaviour of semiconductors are decided by two fundamental properties namely energy gap and refractive index. In general, the threshold of photon absorption of a semiconductor determines the energy gap whereas refractive index is a measure of transparency to the incident photon. The correlation between these two optoelectronic parameters of semiconductors has remained a subject of intensive research in recent times because of its role in semiconductor band structures. Moreover, electronic properties such as atomic polarizability and dielectric constant depend on the refractive of the materials which ultimately can be calculated from the knowledge of the energy gap.

Refractive index of a material is known to decrease with energy gap and therefore, these two fundamental quantities of a material are believed to have certain correlation. Over a period of time there have been many attempts to find a suitable relationship, both empirical and semi empirical, between the energy gap and refractive index of semiconductors [3-21]. Many of the empirical relations relate the refractive index $n$ to the energy gap $E_g$ directly whereas some relations are proposed to calculate $E_g$ from electronegativity first and then from the calculated $E_g$, $n$ is determined. Moss \cite{Moss50, Moss52, Moss85} on the basis of photoconductivity showed that the electron energy levels are scaled down by a factor of $\frac{1}{\epsilon^2_{eff}}$, where $\epsilon_{eff}$ is the effective dielectric constant as felt by the electron in the material. The effective dielectric constant is approximately equal to the square of the refractive index of the material. Penn  proposed a simple model for isotropic systems with reasonable applications to a liquid or amorphous semiconductors \cite{Penn62}. Ravindra and his collaborators have proposed an empirical relation for refractive index for semiconductors which is linear in energy gap  by assuming that the valence and conduction bands are more or less parallel to each other along the symmetry directions \cite{Ravi79, Gupta80}. The Ravindra relation was conceived to be an approximation of the Penn model. Several other modifications of the Ravindra relations were also proposed to get different properties of many infrared materials \cite{Reddy92, Reddy95}. Based on the oscillatory theroy and assuming the UV resonance energy has a constant difference with energy gap, Herve and Vandamme proposed a theoretical relation for refractive index which they claimed to provide good results with lowest deviation for III-V, I-VII and chalcopyrites \cite{Herve94, Herve95}. Reviews on the energy gap and refractive index of semiconductors based on these models can be found at \cite{Moss85,Ravi07, Ravi81}. Gopal in attempt to modify the Penn model for high frequency dielectric constant obtained a relation between these two fundamental properties of semiconductors \cite{Gopal82}. Later, Reddy and his collaborators investigated about the correlation among various properties of semiconductors and proposed some empirical relations for different compounds \cite{Reddy08, Reddy09, Reddy03}. Anani et al. have proposed a formulation for refractive index of III-V semiconductors \cite{Anani08}. Recently, Kumar and Singh proposed an empirical relation for refractive index described through a power law behaviour on the energy gap by fitting the model parameters to a number of experimental energy gap and refractive index data \cite{Kumar10}. In that work, they have claimed that, the proposed model can be equally applied for the determination of optoelectronic properties for different semiconductors including ternary chalcopyrites.

It is needless to emphasize that, a proper design of optoelectronic device requires a detailed knowledge of the refractive indices of materials and hence an accurate and reliable energy gap-refractive index relation is indispensable. From experimental point of view, energy gap of semiconductors are available for a wide range of materials. However, refractive index data for different group of semiconductors including narrow band gap region are not available to a satisfactory extent \cite{Paskov97}. This necessitates the formulation of both theoretical and empirical relationship between these two fundamental parameters.

The purpose of the present work is to study some of the well known energy gap- refractive index relations in the context of elemental and binary semiconductors.  The choice of the empirical relations is very much personal one and is on the basis of their frequent use and discussion in literature. It is obvious that, the already proposed relations are region specific. From a set of tabulated experimental values of energy gap and refractive index of semiconductors and alkali halides, we propose a relation which can be applied equally well to all possible regions of energy gap. The organisation of the present work is as follows. In sect-2, we have reviewed the energy gap- refractive index relations and on the basis of a fit to the experimental data spanning a wide region of energy gap, we propose a new relation for the refractive index. In sect-3, we have compared our results with the known values for different group of semiconductors and alkali halides. Also, we have applied the empirical model for the calculation of refractive indices for ternary compounds. The temperature variation of refractive index for different semiconductors are also discussed. In sect-4, we have extrapolated our results to ternary compounds. At the end, we summarize our results.

\section{Basic formulations}

The correlation of refractive index and energy gap in semiconductor has been a subject of intensive research interest for a long time and started with the semi empirical relation of Moss as early as in 1950. Following Moss relation, there have been many similar relations have been proposed for the calculation of refractive index of materials. In the present section, we review and analyse some of the well known energy gap-refractive index relations as available in literature. These relations are widely used to calculate the refractive index of different group of semiconductors and alkali halides. 

\vspace{0.3cm}
Moss relation:
\begin{equation}
n^4E_g=95 eV,
\end{equation}

where $n$ and $E_g$ are respectively the refractive index and energy gap.

\vspace{0.3cm}
Ravindra relation:
\begin{equation}
n=4.084-0.62E_g.
\end{equation}
\vspace{0.3cm}

Herve-Vandamme relation:
\begin{equation}
n^2=1+\left(\frac{A}{E_g+B}\right)^2,
\end{equation}
where $A$ is the hydrogen ionization energy $13.6eV$ and $B=3.47eV$ is a constant assumed to be the difference between the UV resonance energy and band gap energy.
 
\vspace{0.3cm}
Reddy relation:
\begin{equation}
n^4(E_g-0.365)=154.
\end{equation}

\vspace{0.3cm}
Kumar and Singh relation:
\begin{equation}
n=KE_g^C,
\end{equation}
where $K=3.3668$ and $C=-0.32234$.

\begin{figure}[h!]
\begin{center}
\includegraphics[width=1\textwidth]{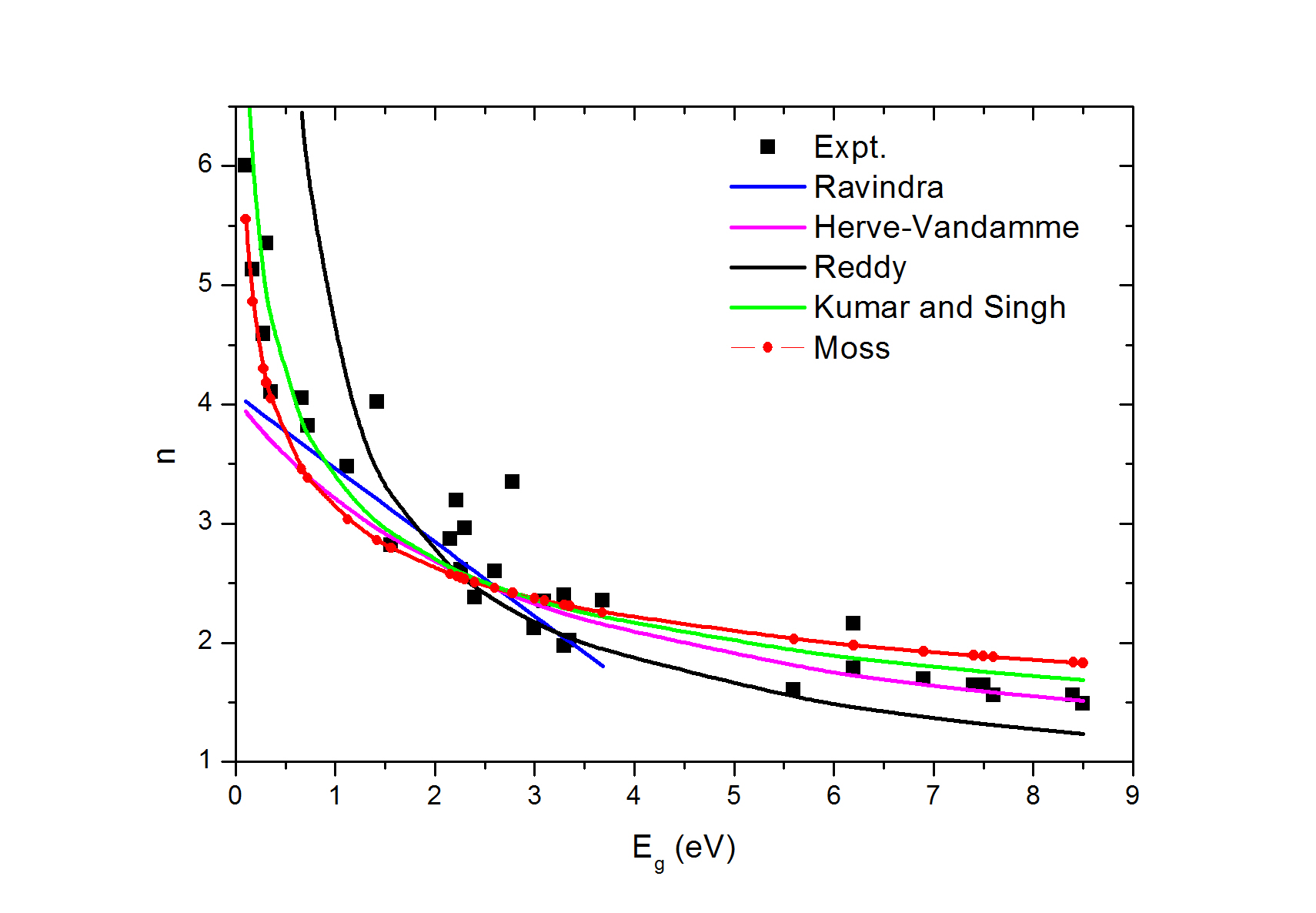}
\caption{Refractive index of semiconductors as function of energy gap. The experimental values are shown as unconnected solid black squares. Calculated refractive indices from some well known relations are also shown in the plot.}
\end{center}
\end{figure}

\begin{figure}[h!]
\begin{center}
\includegraphics[width=1\textwidth]{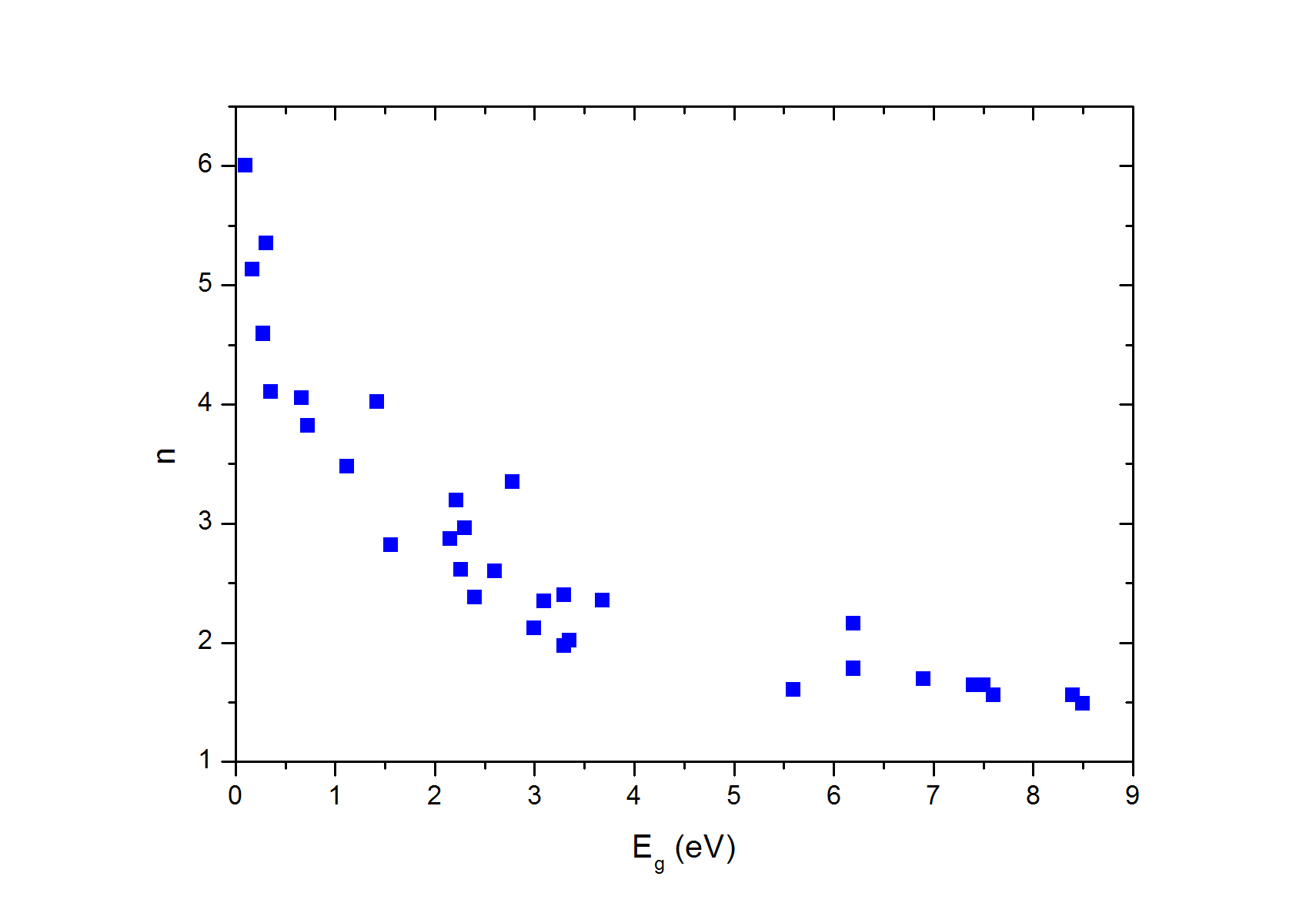}
\caption{Refractive index of semiconductors as function of energy gap. The experimental values of the refractive index corresponding to the respective experimental values of energy gap of some elemental and binary semiconductors for a wide range of energy gap are shown in the plot. The experimental values are collected from different reliable sources \cite{Kumar10, Weber03, Kasap07, Lide99}.}
\end{center}
\end{figure}

The above mentioned relations have been proposed with claims to have good agreement with experimental values. In Figure-1, we have plotted these relations for refractive index as function of energy gap. Also, in the figure, the known or experimental values of refractive index of some elemental and binary semiconductors including alkali halides are shown against their experimental energy gap. The experimental values have been collected from different sources \cite{Kumar10, Weber03, Kasap07, Lide99}. One can note from the figure that, except the Ravindra relation all other relations show a common trend. Ravindra relation being a linear one has its own shortcomings. Eventhough this relation works well in a fairly low and medium values of energy gap, it fails in the very low and high energy gap region. Further, the refractive index becomes negative for Ravindra relation for semiconductors having energy gap greater than $6.587 eV$ and hence cannot be used for materials with refractive index greater than 4.1. In other words for materials like GeTe, InSb, PbSe, PbSnTe and PbTe with larger energy gap ($E_g> 6.587 eV$) this equation may not be valid.  

Herve-Vandamme relation has been derived from the oscillatory theory assuming a constant difference of energy between band gap and UV resonance energy which can be thought of as a modified version of Moss relation. The behaviour of Herve-Vandamme and Moss relation are similar from a low value of energy gap i.e $E_g\simeq 1 eV$ to large values. In their work, Herve and Vandamme claimed that, their model provides a good fit for III-V, I-VII and chalcopyrites. For materials with high band gap energy, this model is somewhat accurate. However, at low values of energy gap ($E_g<1.4 eV$) this relation fails to predict the experimental refractive index. 

Moss relation fairly follows the trend of the experimentally determined refractive index. Moss relation is restricted by the structure of materials and at very low values and at large values of energy gap Moss relation is not a good one to rely upon for the calculation of  refractive index of materials. It can be observed from the figure that, Moss relation is reliable in the range $0.17 eV<E_g<3.68 eV$ of the energy gap.

Reddy relation is a modification of Moss relation with a constant term subtracted from the energy gap and has the same behaviour as that of Moss relation. It is certainly a bit improved than Moss'. This relation gives a better agreement to the experimental values in the medium to reasonably high range of energy gap i.e $1.1 eV <E_g<6.2 eV$. However, this relation is invalid for materials with band gap less than $0.365 eV$ and cannot be used for infra red materials like lead salts and InSb. Also, for very large values of energy gap, Reddy relation substantially deviates from experiments.

Kumar and Singh have tabulated the experimental values of refractive index corresponding to the energy gap for good number of elemental  and binary semiconductors. They have followed a Moss type relationship and fitted the tabulated values to the relation empirically. Therefore, it is obvious that, their relation follows the same trend as that of Moss' and Reddy's but with a stiffer slope at low values of energy gap. At lower energy gap region, this relation remains in between the Moss relation and Reddy relation but in the higher energy gap region, it remains above the Moss but below the Reddy's. In very low region of energy gap, Kumar-Singh relation over predicts the value of refractive index where as in the medium range it under predicts. Of course, one can note from the figure that, it succeeds in predicting the refractive index in the range $2 eV<E_g<4 eV$ and a bit fairly at higher range i.e $6.2 eV<E_g<8.5 eV$.

Besides these relations as mentioned above, some other relations have been proposed where, the refractive index is calculated as a function of electro negativity of materials. All those relations have their success and failures and have been discussed in literature elsewhere. In the analysis of all of these relationship between the refractive index and energy gap, one striking feature we observe that, most of the relations are in good agreement with the experimental values in the medium gap region in  the range $2eV-4eV$. Besides this, some agree well in the low energy gap region and some in higher gap region. In order to get a clear view of the experimental trend of the refractive index with respect to the energy gap, in Figure-2, we have plotted the known refractive index values of all the known elemental and binary semiconductors considered in this work. It is clear from figure-2 that, the experimental refractive index is in general decreases exponentially with energy gap. In view of these facts, in the present work, we have fitted an exponential empirical formula to these data of experimental refractive index and energy gap of some elemental and binary semiconductors of a wide range of energy gap ranging from low value of $E_g=0.1 eV$ to a reasonably high value, $E_g=8.5 eV$. The experimental values were taken from some well known resources. The proposed relationship for those data is

\begin{equation}
n=n_0+n_1e^{-\frac{E_g}{\mu}}\label{f1}.
\end{equation}
The parameters of the above relation for the best fit are found to be $n_0=1.65752 \pm 0.14605 $, $n_1=3.78368\pm 0.21302$, and $\mu=1.85447 \pm 0.25777 eV$. For the fit of the experimental values, the reduced $\chi^2$ is found to be 0.1345 and the adj $\chi^2$ is found to be 0.904. However, in order to provide the formula to a better shape, we rewrite the equation as

\begin{equation}
n=n_0\left[1+\alpha e^{-\beta E_g}\right]\label{m1}.
\end{equation}

\begin{figure}[h!]
\begin{center}
\includegraphics[width=1\textwidth]{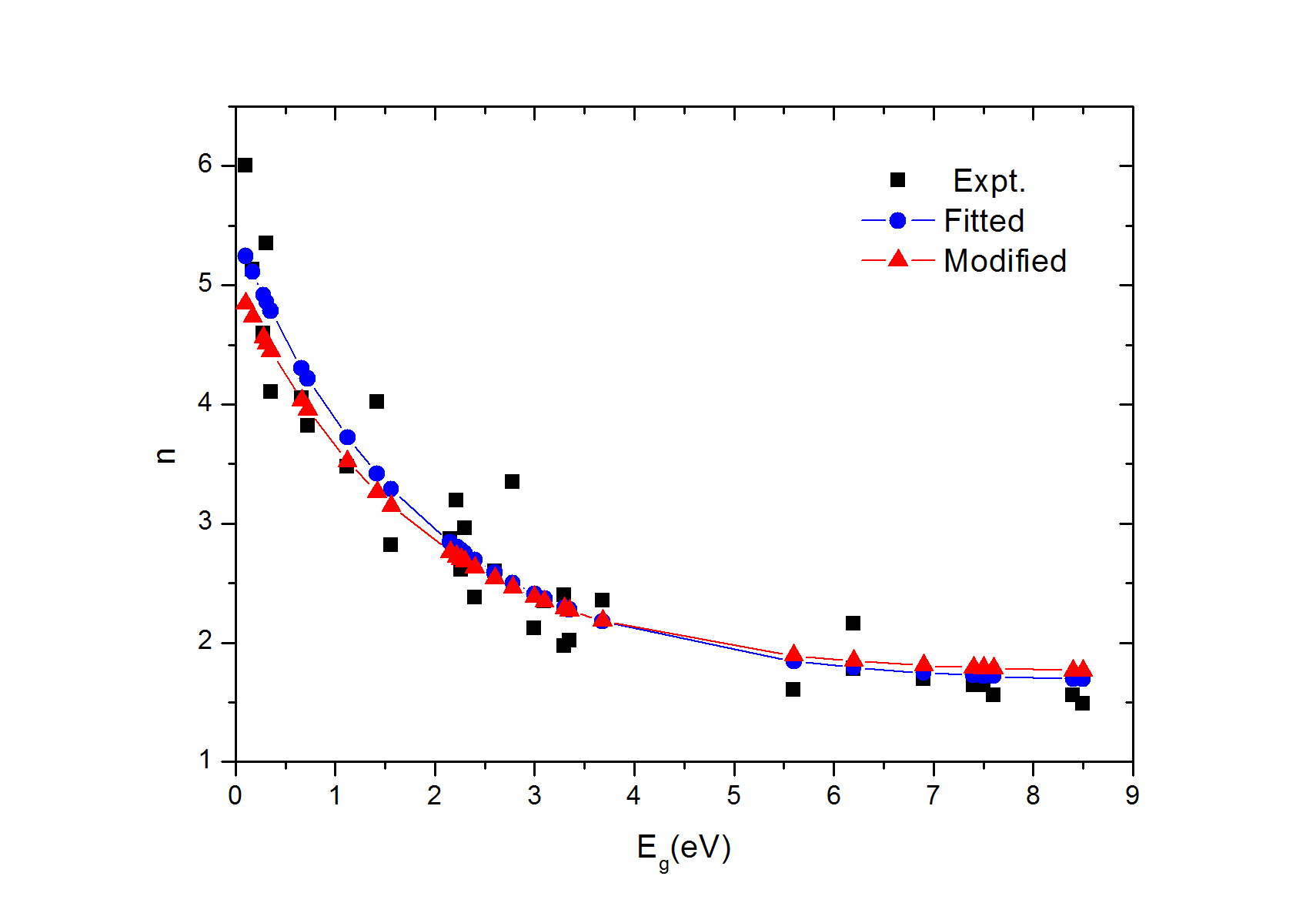}
\caption{Refractive index as function of energy gap as calculated using the fitted relation (eq. \eqref{f1}) and modified relation (eq. \eqref{m1}) of the resent work. The experimental values are shown as unconnected solid black squares for comparison.}
\end{center}
\end{figure}

The new parameters $\alpha, \beta$ and $n_0$ appearing in the above equation are again adjusted so as to get good agreement with the experimental values  of elemental and binary semiconductors over a wide range of energy gap. The parameters of the modified relation are $n_0=1.73 $, $\alpha=1.9017$ and $\beta=0.539 (eV)^{-1}$. While modifying the fitted relation as in eq. \eqref{f1}, we have kept intact the general behaviour of the model by retaining the exponential term. Only the first term in eq. \eqref{f1} has been readjusted with certain normalisation so as to get a better shape and results. The calculated refractive indices over wide range of energy gap for the fitted and the modified relations as in equations \eqref{f1} and \eqref{m1} are plotted in Figure-3. In order to compare our results with that of the known ones, the experimental values are also shown in the figure. In the region of very low values of energy gap i.e. $E_g <0.17 eV$, there occurs a little discrepancy between our results and the experimental values. In the medium range of energy gap i.e. $E_g =2 eV$ to $6 eV$, the agreement with the known value is excellent. At higher side of the energy gap, our calculated values remain slightly above the experimental values. Overall, both the relations agree well with the known values over a wide range of energy gap.

\section{Applications to Binary Semiconductors}

We have used the proposed relation with the modified form for the prediction of refractive index for binary semiconductors and alkali halides separately and compared our results with the results calculated from the well known formulations. As can be seen from different plots, our results are more consistent with the known values than that of other calculations.

\subsection{II-VI group}
Wide band gap II-VI semiconductors have potential applications in optoelectronic devices such as LEDs and LDs operating in blue-green spectral range \cite{Kasap07}. ZnS, ZnSe and ZnTe find applications as blue lasing materials and can be used in the fabrication of optical wave guides and modulated heterostructures \cite{Khenata06,Hasse91, Tamargo91}. Oxides in this group of semiconductors like ZnO are used in technological applications such as nanomedicines \cite{Grone06} and optoelectonic devices \cite{Murt12}. Extensive theoretical investigations of the structural ability and electronic properties of ZnO from Density Functional Theory (DFT) have been made in literature \cite{Wong13, Wong13a}. Recently there have been a lot of studies on the elastic, electronic and optical properties of some II-VI group semiconductors \cite{Khenata06, Reshak06, Reshak07, AlDouri10, Abdul12, Trojnar12, Lippens91, Lefeb94, Umar12, Reshak14}. For a review on the optical properties of II-VI semiconductors one can refer to \cite{Reynolds65}. In Figure-4, the calculated refractive indices for II-VI group binary semiconductors such as ZnS, ZnO, ZnSe, CdS, CdTe, SrTe, BaO, MgSe  etc. are plotted as function of energy gap. For comparison, calculations of refractive index using the prescriptions due to Moss, Herve-Vandamme, Ravindra and Reddy et al. are given. The known values of refractive index for II-VI group of semiconductors are also shown in the figure. As earlier, relations due to Reddy et al., Moss, Herve and Vandamme show the experimental trend. Ravindra relation being a linear one passes through the experimental points in the region $1.5 eV <E_g < 3.5 eV$ and deviates uch in the higher energy gap region. Reddy relation goes much above the experimental points and hence poorly fits this particular group of semiconductors. Baring some points, Herve and Vandamme relation predicts refractive index close to known values in the region $2 eV <E_g < 4 eV$. But in general, the trend for this relation is below the known ones. Moss relation passes almost in the middle through the known points in the region $2 eV <E_g < 5.5 eV$. The calculations from the proposed relation in the present work eq. \eqref{m1} pass in between the Herve-Vandamme and Moss relations. Even though it under predicts the refractive index in the mid energy gap region, but as a whole, the number of points at which the prediction agrees with known values are more as compared to other calculations. Also, in the low and high energy gap regions, the predictions are much better than others.

\begin{figure}[h!]
\begin{center}
\includegraphics[width=1\textwidth]{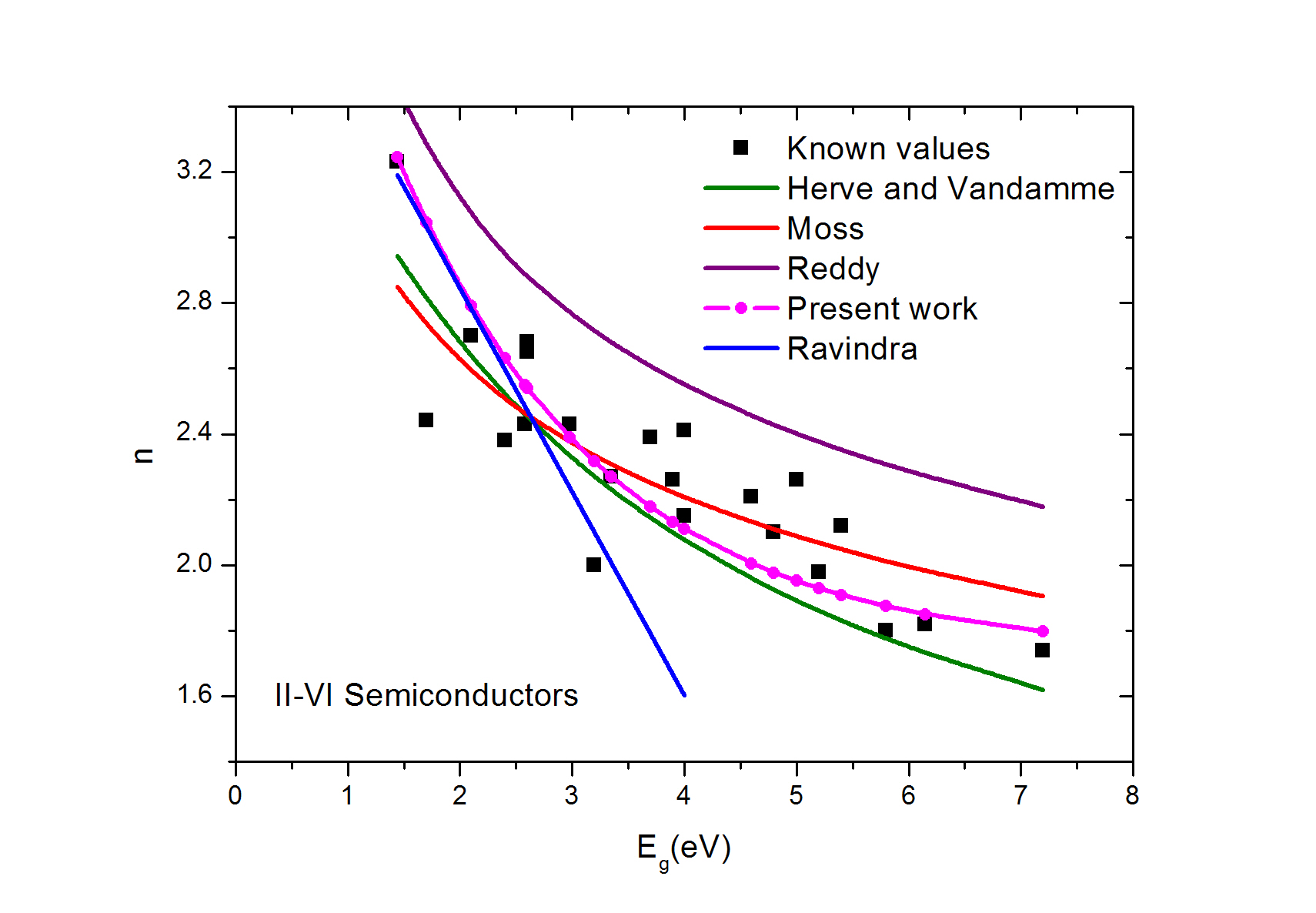}
\caption{Refractive index of II-VI group semiconductors as function of energy gap using our modified relation eq.(7). The known values and calculations from some well known relations are also shown for comparison.}
\end{center}
\end{figure}

\subsection{III-V group}
III-V group of semiconductors have interesting applications in optoelectronics and photovoltaics because of their direct band gaps and high refractive indices. They find wide applications in the fabrications of high efficiency solar cells. Large breakdown fields, high thermal conductivities and electron transport properties of III-V nitrides such as GaN, InN, AlN make them suitable for novel optoelectronic applications in visible and ultra violet spectral range \cite{Kasap07}. In recent times there have been a lot of interest in the calculation of the electronic and optical properties of III-V binary semiconductors and their alloys \cite{Johnson90,Reshak06a,AlDouri11, Bredin94, Borak05, Levine91, Reshak07a, Reshak05,Tit10,Haq14}. In Figure-5, we have plotted the calculations for refractive index for some known III-V group binary semiconductors using the proposed relation in eq. (7). Also, we have shown the general trend of some other relations discussed earlier in the work. The known values are shown as black solid squares in the figure. More or less all the plotted relations follow the experimental trend of the refractive index of this group of semiconductors. Ravindra relation follows the trend  all through the region shown, but it disagree with the prediction of experimental value. Moss relation fails for this particular group of semiconductors. As in the previous case of II-VI group of semiconductors, Reddy relation goes above the acceptable trend. However, at large energy gap, Reddy relation predicts the refractive index well at least for AlP and AlAs with refractive indices 2.76 and 2.96 respectively. Interestingly, for the present group of semiconductors, our calculation with the proposed modified relation follows closely the experimental trend, although at low values of energy gap the agreement is poor. However, with our relation, the refractive indices of GaSb, InP, GaAs, AlSb and GaN are predicted much closer to known values. The refractive indices for these semiconductors are calculated to be 3.86, 3.4,3.28,3.195 and 2.26 respectively in our modified relation in eq.(7).

\begin{figure}[h!]
\begin{center}
\includegraphics[width=1\textwidth]{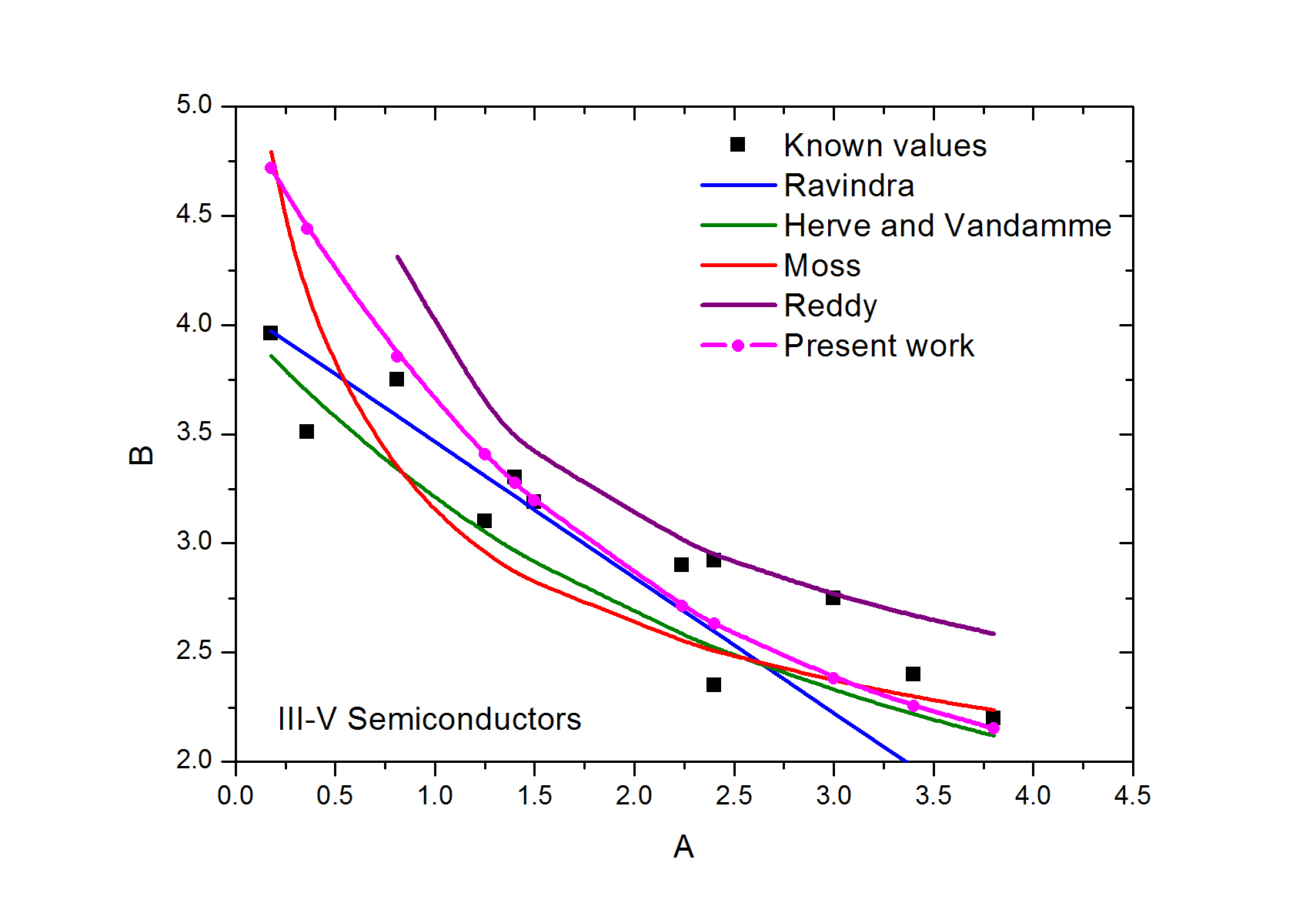}
\caption{Refractive index of III-V group semiconductors as function of energy gap using our modified relation in eq.(7). The known values and calculations from some well known relations are also shown for comparison.}
\end{center}
\end{figure}

\subsection{Alakli Halides}
The refractive indices of a good number of alkali halides as calculated from our proposed relation are plotted as function of energy gap in Figure-6. The results of well known relations and the known values are also shown in the figure for comparison. Except Herve-Vandamme relation, all other relations including ours are not in well agreement with the known values. It is clear from the figure that, Moss relation and Reddy relation completely fail to predict the refractive indices of alkali halides. Ravindra relation , even though not able to predict correctly for alkali halides, is bit closer to the known values in the low energy gap region within the range $E_g=2.4-3.1 eV$. Since the energy gap of alkali halides are much larger than $6.587 eV$, the relation quickly goes for negative indices. One can observe from the figure that, Herve and Vandamme relation provides an excellent fit for the refractive indices of alkali halides over a wide range energy gap. At very high energy gap region, this relation predicts a bit higher values than the known ones. In the low energy gap region, also the agreement with known values are good. Our fitted relation is a bit closer to the known values for CsI, LiBr, NaI, KI, CsBr, LiCl, RbI, NaBr and CsCl. However, the modified relation predict a slightly higher values for these alkali halides. Our relations being empirical ones and have been fitted keeping an eye on a list of tabulated elemental and binary semiconductors may not be that much valid for alkali halides. But one can note that, our fitted as well as the modified relations follow the same trend as that of the known values through out the band gap space considered in this work. The reason may be inherent in the relation itself and may it need a further theoretical investigation and justification. Since the experimental trend is maintained by our relation, we wished to keep the decremental behaviour part of our empirical relation intact and readjusted the normalisation constant $n_0$ so as to fit for alkali halides. The renormalised constant for alkali halides is considered to be $n_0=1.57 eV$. The readjusted relation now able to predict to an acceptable accuracy for the refractive indices of CuI, AgI, AgBr, CuBr, CuCL, LiI, CsI,LiBr, NaI, KI, CsBr,LiCL, RbI, NaBr, CsCl, KBr, RbBr, NaCl. However, for large value of energy gap,  the readjusted relation predicts higher than the known values but can be more reliable compared to Reddy, Moss and Ravindra relations.

\begin{figure}[h!]
\begin{center}
\includegraphics[width=1\textwidth]{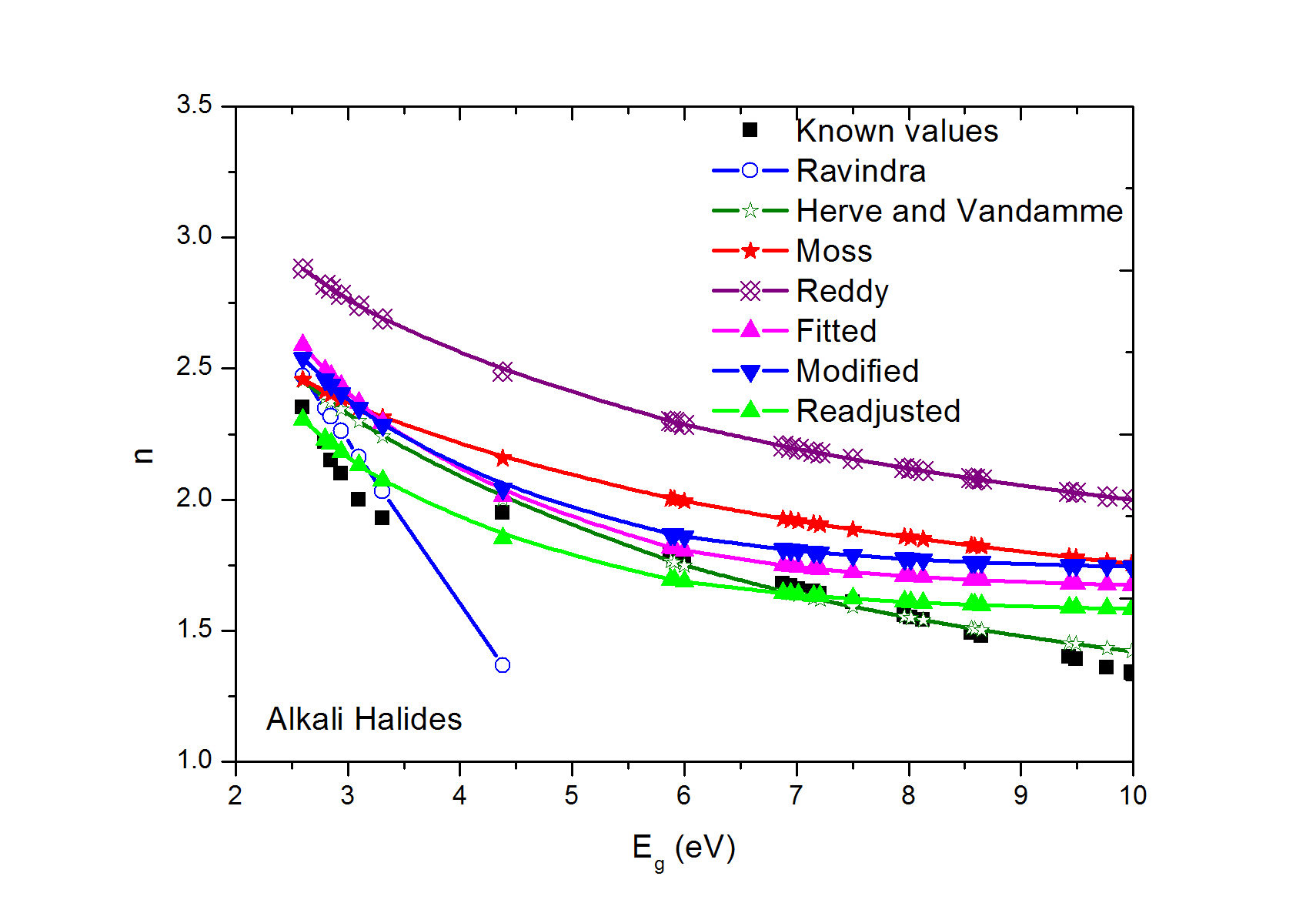}
\caption{Refractive index of some Alkali Halides as function of energy gap using our fitted (eq. (6)) and modified relations (eq.(7)). We have readjusted the normalisation constant parameter to fit the data for alkali halides. The known values and calculations from some well known relations are also shown for comparison.}
\end{center}
\end{figure}

\subsection{Dielectric constants of some III-V group semiconductors}
Semiconductors fabricated from elements of group III i.e Al, Ga, In and group V i.e P, As, Sb have much importance in the sense that they are used in optoelectronic devices. These compounds exhibit zinc-blende crystal structure. The success of the proposed refractive index-energy gap relation in exponential form can also be tested for determination of the dielectric constants of some III-V group binary semiconductors. The dielectric constant of a material is related to the refractive index as $\epsilon_ r=n^2$. The refractive index can be calculated from the exponential relation as proposed in the present work and consequently dielectric constant can be calculated. The calculated dielectric constants of some of III-V materials are given in table-1. The experimental values of dielectric constants of the materials are also given in the table. For comparison, the values calculated with Herve and Vandamme relation are shown in the table. The calculated dielectric constants with our fitted and modified relations are in good agreement with the experimental values. Of course for some materials like InSb, InAs and AlP our results do not agree with the experimental values. This may be due to the over simplified expression  $\epsilon_ r=n^2$ used in the calculation or due to some inherent physical processes which have not been properly looked into during the calculation.

\begin{table}
\caption{Dielectric constants of some III-V group semiconductors.}
\centering
\begin{tabular}{l|c|c|c|c}
\hline \hline
Compounds	&	Modified 	&	Fitted 	 	& Expt. &	Herve and\\
			&	relation	&	relation	&		&	 Vandamme\\
\hline
$InSb$		&	22.24			&	25.92 	& 15.7	&	14.88\\
$InAs$		&	19.71			&	22.79 	& 12.3	&	13.61\\
$GaSb$		&	14.87			&	16.83 	& 13.7	&	11.09\\
$InP$			&	11.61			&	12.86 	& 9.6	&	9.30\\
$GaAs$		&	10.74			&	11.81 	& 10.9	&	8.798\\
$AlSb$		&	10.21			&	11.17	& 10.2	&	8.487\\
$GaP$			&	7.36			&	7.77 	& 8.5	&	6.67\\
$AlAs$		&	6.93			&	7.26	& 10.2	&	6.367 \\
$InN$			&	6.93			&	7.26	& 5.5	&	6.367\\
$AlP$			&	5.68			&	5.8		& 8.5	&	5.418\\
$GaN$			&	5.1				&	5.12	& 5.18	&	4.918\\
$AlN$			&	4.64			&	4.60	& 4.8	&	4.499\\
\hline
\end{tabular}
\end{table}

\subsection{Temperature  variation of Refractive index}

Refractive index of semiconductors are known to vary with temperature. In fact, the refractive index increases with the increase in temperature. By differentiating eq. \eqref{m1} with respect to temperature $T$, we get

\begin{equation}
\frac{dn}{dT}=-(n-n_0)\beta \frac{dE_g}{dT},
\end{equation}
where we have assumed that the constant parameters appearing in \eqref{m1} are temperature independent. Now the relative variation of refractive index with temperature can be expressed for \eqref{m1} as

\begin{equation}
\frac{1}{n}\frac{dn}{dT}=-\left(1-\frac{n_0}{n}\right)\beta \frac{dE_g}{dT}.
\end{equation}

We have calculated the temperature variation of refractive index for different binary semiconductors using the known temperature variation in energy gaps $\frac{dE_g}{dT}$. The calculated values are given in table-2. For comparison, in the table, we have also given the values calculated using Herve and Vandamme relation taken from \cite{Ravi07}. Our results using the modified relation for refractive index provide reasonable values for the temperature variation of refractive index but do not agree in toto. One possible reason behind the disagreement between the calculated  values and that of Herve and Vandamme relation may be  that, the constant parameters appearing in eq. \eqref{m1} should be temperature dependent. In their calculation, Herve and Vandamme have taken the parameter $B$ in eq. (3) to be temperature dependent. The slope of the parameter $B$ with respect to temperature $T$ in Herve and Vandamme(HV) relation is taken to be $\frac{dB}{dT}\simeq 2.5 \times 10^{-5} eV/K$. Similarly, in our relation if at least, one parameter say $\alpha$ be made temperature dependent then, we hope that, the discrepancy can be removed. Considering a temperature dependence of $\alpha$, the modified relation \eqref{m1} can be rewritten as

\begin{equation}
n(T)=n_0\left[1+\alpha(T) e^{-\beta E_g(T)}\right]
\end{equation}
and consequently, the relative temperature variation of refractive index can be 

\begin{equation}
\frac{1}{n}\frac{dn}{dT}=\left(1-\frac{n_0}{n}\right)\left[\frac{1}{\alpha} \frac{d\alpha}{dT}-\beta \frac{dE_g}{dT}\right].
\end{equation}

The relative temperature variation of refractive index depends on the temperature variation of the energy gap and the relative temperature variation of the parameter $\alpha$. If for any group of semiconductors or an individual semiconductor, this factor is known, then, $\frac{1}{n}\frac{dn}{dT}$ can be easily calculated. However, in the present study, we limit ourselves only to the constant nature of $\alpha$ and defer the temperature dependence of the parameters appearing in \eqref{m1} for future investigation.

\begin{table}
\caption{Temperature variation of refractive index of semiconductors}
\centering
\begin{tabular}{l|c|c|c|c|c}
\hline \hline
Compounds	&$E_g$	& $n$	&$\frac{dE_g}{dT}$	&	$\frac{1}{n}\frac{dn}{dT}$ 	& $\frac{1}{n}\frac{dn}{dT}$ \\
			&$(eV)$	& 		& $(meV K^{-1})$	&	 $(\times 10^{-6}K^{-1})$	&$(\times 10^{-6}K^{-1})$\\
			&		& eq.(7)&					&	present work				& HV\\
\hline
$InSb$		&0.18	&	4.72&	-280			&	95.55						&	69\\
$PbSe$		&0.278	&	4.56&	510				&	-170.65						&	-210\\
$Ge$		&0.67	&	4.02&	-370			&	113.66						&	69\\
$GaSb$		&0.75	&	3.93&	-370			&	111.55						&	82\\
$Si$		&1.1	&	3.55&	-280			&	77.34						&	40\\
$InP$		&1.35	&	3.32&	290				&	-74.84						&	27\\
$GaAs$		&1.43	&	3.25&	-390			&	98.39						&	45\\
$AlAs$		&2.15	&	2.76&	-400			&	80.58						&	46\\
$AlP$		&2.41	&	2.63&	-370			&	68.12						&	36\\
$SiC$		&2.86	&	2.43&	330				&	-51.46						&	29\\
$GaN$		&3.5	&	2.23&	-480			&	57.89						&	26\\
$C$			&5.48	&	1.90&	-50				&	2.43						&	4\\
\hline
\end{tabular}
\end{table}

\section{Applications to ternary compounds}

Ternary compounds are highly promising materials and are used in high efficiency solar cells and quantum electronic devices \cite{Bodnar11, Bodnar07}. Ternary chalcopyrites have attracted a lot of attention in recent times both from the theoretical and experimental point of view of investigation \cite{Tit10a, Khan14, Ouahrani10, McCann95, Ghosh14, Reshak05a, Reshak08}. In view of the recent interest and wide applications of ternary compounds, we have used the proposed empirical relation to calculate the refractive index of some ternary compounds. The energy gaps of the ternary compounds are calculated from the Duffy formulation \cite{Duffy80, Duffy90} using the electronegativity of the materials i.e $E_g=3.72 \Delta x$, where, $\Delta x$ is the electronegativity of the material. The calculated values of the refractive indices are given in table-3. In the table the values calculated by using HV relation are also given for comparison. The calculated results provide some reasonable estimates of refractive index of ternary compounds.

\begin{table}
\caption{Refractive index of some ternary semiconductors.}
\centering
\begin{tabular}{l|c|c|c|c|c}
\hline \hline
Compounds	&$\Delta x$& $E_g$	&	Modified 	&	Fitted for  	& HV \\
			&$eV$	& ($eV$)(Duffy)	&	relation	&	Alkali Halides	&\\
\hline
$CuAlS_2$	&0.938		&	3.49	&2.23	&	2.03	&2.19\\
$CuAlSe_2$	&0.723		&	2.69	&2.50	&	2.27	&	2.42\\
$CuAlTe_2$	&0.552		&	2.05	&2.82	&	2.56	&	2.66\\
$CuGaS_2$	&0.643		&	2.39	&2.64	&	2.39	&	2.53\\
$CuGaSe_2$	&0.455		&	1.69	&3.05	&	2.77	&	2.82\\
$CuGaTe_2$	&0.268		&	0.997	&3.65	&	3.31	&	3.20\\
$CuInS_2$	&0.403		&	1.499	&3.19	&	2.90	&	2.91\\
$CuInSe_2$	&0.279		&	1.038	&3.61	&	3.28	&	3.18\\
$CuInTe_2$ &0.254		&	0.95	&3.71	&	3.36	&	3.24\\
$AgAlSe_2$ &0.683		&	2.54	&2.57	&	2.33	&	2.47\\
$AgAlTe_2$	&0.608&	2.26&	2.70&	2.45&	2.57\\
$AgGaS_2$	&0.723&	2.69&	2.50&	2.27&	2.42\\
$AgGaSe_2$&	0.482&	1.79&	2.98&	2.71&	2.77\\
$AgGaTe_2$&	0.294&	1.09&	3.55&	3.23&	3.14\\
$AgInS_2$&	0.534&	1.99&	2.86&	2.59	&2.69\\
$AgInSe_2$&	0.332&	1.23&	3.42&	3.10&	3.06\\
$AgInTe_2$&	0.268&	0.99&	3.65&	3.31&	3.20\\
$ZnSiP_2$&	0.562&	2.09&	2.79&	2.54&	2.64\\
$ZnGeP_2$&	0.533&	1.98&	2.86&	2.59&	2.69\\
$ZnSnP_2$&	0.444&	1.65&	3.08	&2.79&	2.84\\
$ZnSiAs_2$&	0.456&	1.69&	3.05&	2.77&	2.82\\
$ZnGeAs_2$&	0.308&	1.15&	3.50&	3.18	&3.11\\
$ZnSnAs_2$&	0.269&	1.00&	3.65&	3.31&	3.20\\
$CdSiP_2$&	0.656&	2.44&	2.61&	2.37&	2.51\\
$CdGeP_2$&	0.461&	1.71&	3.03&	2.75&	2.81\\
$CdSnP_2$&	0.313&	1.16&	3.49&	3.16&	3.10\\
$CdSiAs_2$&	0.415&	1.54&	3.167&	2.87	&2.89\\
$CdGeAs_2$&	0.164&	0.61&	4.10&	3.72&	3.48\\
$CdSnAs_2$&	0.069&	0.26&	4.59&	4.17&	3.78\\
\hline
\end{tabular}
\end{table}

\section{Summary and Conclusion}
In the present work, we have studied the refractive index of semiconductors as function of energy gap. Refractive index and energy gap are the two important quantities that decide the optical and electronic behaviour of semiconductors used for optoelectronic applications. Refractive index is known to decrease with energy gap. There are some well known empirical and semi empirical relations for calculation of refractive index from energy gap. In the present study we have analysed all those well known relations both from the point of view of their success and failures. Also, we have proposed an empirical relation based upon the known energy gap data corresponding to their refractive index of some elemental and binary semiconductors over a wide range of energy gap. Keeping an eye on the general trend of the data, an exponentially decreasing refractive index is proposed and the parameters are fitted to get least deviation. The proposed relation is then separately applied to different groups of semiconductors such as II-VI, III-V and Alkali Halides. Our calculations are also compared with the calculations of other works. Except for alkali halides, our formulation is found to be in good agreement with the known values compared to others. For alkali halides, except Herve and Vandamme relation, no other relations considered in the work are in good agreement over the range of energy gap  considered. Even if our relations, both the fitted and modified ones, follow the same trend as that of the known ones,  there remained some disagreement between the known values and calculated values. In order to bridge the gap, we readjusted the parameters keeping the trend intact to get a satisfactory agreement with the known values. The readjusted relation could be able to predict the refractive indices of a good number of alkali halides. In order to test our relation, we have also applied it to calculate the dielectric constant of some III-V semiconductors to get very good results. The proposed relation is also used to calculate the refractive index of some ternary semiconductors and the temperature variation of refractive index of some binary semiconductors. Our relation provides  satisfactory results for the temperature variation but do not agree with the calculation of Herve and Vandamme relation. The reason may lie in the temperature dependence of the constant parameters appearing in the proposed relation which should be investigated intensively either for a group of semiconductors or for individual one. Overall, even though, the proposed relation is a simple one, it is found to be successful at least in providing a satisfactory value of refractive index if energy gap is known for a wide range of semiconductors.

\end{document}